\documentclass{article}
\usepackage{amssymb,graphicx,amsmath,url,epstopdf}
\usepackage{geometry,graphics}
\begin{document}

\begin{center}{\large Self Organized Criticality in a two dimensional Cellular Automaton model of a magnetic flux tube with background flow}\\ {\small
B. D\u{a}nil\u{a}\footnote{bogdan.danila22@gmail.com, Astronomical Institute of the Romanian Academy, 15 Cire\c{s}ilor Street, 400487 Cluj-Napoca,  Romania}, T. Harko\footnote{t.harko@ucl.ac.uk, Department of Mathematics, University College London, Gower Street, London
WC1E 6BT, United Kingdom} and G. Mocanu\footnote{gabriela.mocanu@ubbcluj.ro, Astronomical Institute of the Romanian Academy, 15 Cire\c{s}ilor Street, 400487 Cluj-Napoca,  Romania}}\end{center}

\date{}


\begin{abstract}
We investigate the transition to Self Organized Criticality in a two-dimensional model of a flux tube with a background flow.
The magnetic induction equation, represented by a partial differential equation with a stochastic source term, is discretized and implemented on a two dimensional cellular automaton. The energy released by the automaton during one relaxation event is the magnetic energy. As a result of the simulations we obtain the time evolution of the energy release, of the system control parameter, of the event lifetime distribution and of the event size distribution, respectively, and we establish that a Self Organized Critical state is indeed reached by the system. Moreover, energetic initial impulses in the magnetohydrodynamic flow can lead to one dimensional signatures in the magnetic two dimensional system, once the Self Organized Critical regime is established. The applications of the model for the study of Gamma Ray Bursts is briefly considered, and it is shown that some astrophysical parameters of the bursts, like the light curves, the maximum released energy, and the number of peaks in the light curve can be reproduced and explained, at least on a qualitative level, by working in a framework in which the systems settles in a Self Organized Critical state via magnetic reconnection processes in the magnetized Gamma Ray Burst fireball.
\end{abstract}


\section{Introduction}

Many important natural dynamical systems show the presence of long-range spatial and temporal
correlations. For example, spatial scale-invariance is observed in fractal geographical
and topographical structures - mountain ranges, river basins, etc. \cite{Mand}, while temporal scale-invariance in the form of $1/f$-like power spectra is
observed in such diverse phenomena as star flickers and earthquakes \cite{Sor, Aschb, Asch,Wat}. The spatial
scale invariance in particular is in marked contrast to the typical behaviour of
equilibrium thermodynamic systems, where such an invariance can be realized by tuning a parameter
(e.g., the temperature) to a critical value. To describe scale-invariance that occurs
in dynamical systems without the explicit tuning of a parameter, a model pioneered by~\cite{katz1986} and later popularized as Self-Organized Criticality (SOC) by \cite{bak1987} and \cite{bak1988}, has begun to develop.

As a paradigm of self-organized critical behaviour, \cite{bak1987} introduced the sandpile model, which they implement with a Cellular Automaton (CA). A
one-dimensional simple local version of the CA can be briefly described as follows \cite{Kad, Chay}. We start by prescribing two positive integers, $z_c$ and $n$. Then to each site $i = 1 , . . . , L$,
we assign an  integer $h(i)\geq 0$, representing the height of sand at location $i$. The
slope of the sandpile at $i$ is given by $z(i) = h(i) - h(i + 1)$. The system is assumed to
be closed at its left boundary ($i = 0$), and open at the right one ($i = L + 1$). The dynamical process consists in dropping a single
grain of sand onto a randomly chosen site. If the slope at any site $i$ exceeds $z_c$, then $n$ grains of sand fall from $i$
onto the site $i + 1$. This process may cause the slope to exceed $z_c$ at adjacent
sites. If necessary, $n$ grains fall to the right
from any site at which the slope now exceeds $z_c$. The process continues until
all slopes are less than $z_c$. This may be accomplished either with or without the loss
of sand from the right boundary. The entire event caused by dropping a single
grain is called an avalanche. The number of sites from which sand has fallen
is called the size of the avalanche. After an avalanche, a grain is dropped again
onto a randomly chosen site, hence another avalanche occurs, and the process continues indefinitely.
The obtained numerical results \cite{bak1987,Kad} indicate a power law distribution in the sizes of
avalanches. This is exactly the type of distribution found at the critical point of traditional
equilibrium systems. Quantities related to the avalanche sizes also have power law
behaviour. Thereby one can define a set of critical exponents and scaling relations for the system \cite{Chay}.

As an alternative to the numerical study of SOC in cellular automata, one  can
consider models based on differential equations that describe the fluctuations of a conserved
quantity \cite{Hwa,Gar, Grin, Lu1, Lu2,Gil}. More specifically, systems described by diffusion equations
driven by noise terms shows characteristics similar to SOC, with the numerical solutions of partial differential equations on finite grids yielding cellular automata with real-valued states. For example, the basic one dimensional diffusion equation $\partial u/\partial t=(1/2) \partial ^2u/\partial x^2$ can be transformed into the discrete equation $u(x,t+1)=(1/2)\left[u(x+1,t)+u(x-1,t)\right]$ \cite{Sor}. Conversely, the diffusion limits of suitable cellular automaton models yield partial differential equations.   The solutions of such diffusion equations exhibit scale-invariant behaviour, similar to that
observed in sandpiles and related discrete models. It turns out that
SOC can already be understood on the level of the linear stochastic Langevin equation \cite{Chay}.
This is due to the fact that algebraic decay of spatial correlations is a reflection in spatial dimensions of the
algebraic decay in time for systems with conservative dynamics.

The possibility that SOC results in some specific astrophysical phenomena has been intensively investigated recently, and the study of SOC has become a major field of research in astrophysics \cite{Aschb, Asch, flareSOC}. In particular, an interesting model to describe the properties of the accretion disks around black holes was introduced in \cite{Min1, Min2}, where it was suggested that the inner portions of black hole accretion disks may be in a self-organized critical state. Consequently, $1/f$-like X-ray fluctuations are produced, in spite of random mass input. Viscous diffusion processes are also incorporated in the model. Hence in this approach mass accretion occurs either by an avalanche, which is triggered when the mass density of the disk exceeds some critical value, or by gradual gas diffusion. This initial model was further modified in \cite{Tak1}, where a gradual diffusion which occurs regardless of the critical condition was introduced. In \cite{Tak2} it was shown that with only one free model parameter, the cellular automaton model can reproduce some observational data of Cyg X-1.  The relativistic effects were included in the model in \cite{Xiong},
where it was shown that the CA model can produce light-curves and power-spectra for the variability that agree with
the range observed in optical and X-ray studies of AGN and X-ray binaries. It was also pointed out that when general
relativistic effects are incorporated, important differences do appear if the disk
is viewed from directions far from the accretion disk axis. The magnetic activity of an accretion disc using a probabilistic cellular automaton model was simulated in \cite{Pav}. The model is based on three free parameters, the probabilities of spontaneous and stimulated generation of magnetic flux above the surface of the disc, and the probability of diffusive disappearance of flux below the surface. This approach allows steady accretion in a disc by the action of coronal magnetic flux tubes alone, and if the effective viscosity caused by coronal loops is expressed in the usual Shakura-Sunyaev alpha parameter of viscosity, one can obtain numerical values which are in good agreement with observations.  A modified SOC model based on cellular automaton mechanism for producing lognormal flux distribution was presented in \cite{Kun}. In this model the energy released in the avalanche and diffusion in the accretion disk is not entirely emitted instantaneously, with some part of the energy  kept in the disk. Thus the disk increases its energy content, so that the next avalanche will be in higher energy condition,  and more energy will be  released. Hence the later an avalanche occurs, the more amount of energy is emitted from the disk.

On the other hand it was found observationally that a class of Gamma-Ray Bursts (GRBs) called Soft Gamma
Ray Repeaters (SGR) do exhibit SOC characteristics \cite{Asch, Gog}.  From the statistics of the SGR 1806-20 bursts it was shown that the fluence distribution of bursts observed with different instruments is well described by power laws with indices 1.43, 1.76 and 1.67, respectively. Thus, it turns out that the SGR 1806-20 bursts behave in a self-organized critical manner, similarly to earthquakes and solar flares \cite{Gog}. From an astrophysical point of view this behavior suggests that the energy sources for SGR bursts are crust-quakes, due to the evolving, strong magnetic field of the neutron star, rather than any accretion or nuclear reaction processes. A very similar result was obtained for the case of the  X-ray flares of GRBs with known redshifts by \cite{Wang}, who have shown that X-ray flares and solar flares share in common three statistical properties: power-law frequency distributions for energies, durations, and waiting times. All these distributions are specific for the physical framework of a SOC system~\cite{dimitropoulou2011}. Hence we have the interesting result that the statistical properties of X-ray flares of GRBs are similar to solar flares, and therefore both can be attributed to a SOC process. As suggested in \cite{Wang}, both types of flares may be driven by a one dimensional SOC magnetic reconnection process. On the other hand the X-ray flares of GRBs cold be  produced astrophysically in ultra-strongly magnetized millisecond pulsars, or long-term hyperaccreting disks around stellar-mass black holes.

The possibility of the one dimensional SOC modelling of X-ray GRB afterglows,  as well as the possibility of appearance of Self-Organized Criticality  in an one dimensional magnetized flow was carefully investigated in \cite{1dgrb}. A simplified one dimensional grid was used to model the evolution of the magnetized plasma flow. Diffusion laws similar to those used to model magnetic reconnection with Cellular Automata in various astrophysical phenomena were implemented in the model, as well as  a background flow.   Under the assumption that the parameter relevant for X-ray afterglows is the magnetic field, the magnetic energy released by one volume during one individual relaxation event was computed. The obtained results show that indeed in this system SOC is established. The possible applications of this model to the study of the X-ray afterglows of GRBs was also briefly considered.

However, {\it from a strictly theoretical point of view one dimensional magnetic reconnection is not possible.} This raises the interesting and important question of clarifying the relation between GRB observations, showing that indeed one dimensional SOC occurs in the astrophysical process, and their possible relation with one dimensional magnetic reconnection models. In order to answer to this question in the present paper we generalize the one dimensional model introduced  in~\cite{1dgrb}.

Hence in order to describe the temporal evolution of GRBs we develop a two-dimensional CA simulation algorithm, which implements the magnetic induction equation in the MagnetoHydrodynamic (MHD) approximation framework. The bulk advection motion is, as a novelty with respect to other models, taken explicitly into account for various velocity profiles. The end purpose is to find a model which is simple to implement but which still captures the important macroscopic characteristic of observed X-Ray afterglows in GRBs. The main result of our analysis is that even the rigorously implemented physics is 2D, energetic initial impulses in the MHD flow lead to 1D signatures once the SOC regime is established.

The present paper is organized as follows. The simulation procedure is described in Section~\ref{sect1}, and the discretized MHD equation for the magnetic induction are written down.  The simulation results, and some of their astrophysical implications, are presented in Section~\ref{sect2}. We discuss and conclude our findings in Section~\ref{sect3}.

\section[]{Setup of the simulation}\label{sect1}

Although our goal is to keep the presentation in this Section self-consistent, the reader is referred to~\cite{1dgrb} for technical details of the simulation that might be absent from our current presentation. In the following we concentrate on the evolution of the magnetic field in the cosmic environment. From a physical point of view, under certain conditions, the magnetic field topology can change suddenly, thus exhibiting a transition in its behaviour. The transition can be understood and quantified in terms of the Laplacian of the magnetic field achieving a critical value~\cite{dimitropoulou2011}. The evolution equation for the magnetic field is the magnetic induction equation. It describes the time and space evolution of a magnetic field in a non-ideal plasma medium. The magnetic induction equation is derived under the assumption that the MHD approximation holds in the plasma~\cite{Priest2000}, and that there is a background flow. We will present below a cellular automata approach to plasma dynamics. The information about the relevant physical parameter describing the dynamics of the system is stored in the cells of the grid~\cite{CAmodelling}.

\subsection{The magnetic induction equation}

We begin our presentation of the mathematical model to be simulated with the induction equation \cite{Isl, Priest2000}
\begin{equation}\label{magnind}
\frac{{\partial \vec B}}{{\partial t}} = \nabla  \times \left( {\vec v \times \vec B} \right) + \eta {\nabla ^2}\vec B,
\end{equation}
where $\vec{B}=\vec{B}(x,y,z,t)$ is the magnetic field, $\vec{v}=\vec{v}(x,y,z,t)$ is the plasma velocity, and $\eta $ is the total magnetic diffusivity coefficient, $\eta =1/\mu_0 \sigma$, where $\sigma $ is the electrical conductivity, and $\mu _0$ is the magnetic permeability in a vacuum.
We define the control parameter in the plasma flow as
\begin{equation}\vec G =  - \frac{1}{4}{\nabla ^2}\vec B,\end{equation} to be computed for the four neighbours in a two dimensional grid (left, right, up, down).
The configuration we are working with is
\begin{equation}\vec B = (B_x(y,z,t),0,0),\end{equation}
\begin{equation}\vec v = (0,0,v_z(z,t)).\end{equation}
The magnetic field evolution equations then become
\begin{equation}\frac{{\partial B_x}}{{\partial t}} =  - \frac{{\partial (v_zB_x)}}{{\partial z}} + \eta \frac{{{\partial ^2}B_x}}{{\partial {y^2}}} + \eta \frac{{{\partial ^2}B_x}}{{\partial {z^2}}},\end{equation}
\begin{equation} G_x =  - \frac{1}{4}\left( {\frac{{{\partial ^2}B_x}}{{\partial {z^2}}} + \frac{{{\partial ^2}B_x}}{{\partial {y^2}}}} \right).\end{equation}

In the numerical simulations, the time dynamics of the system is usually implemented as one of two different regimes (advection or diffusion), as a function of the value of the control parameter. A connection to the physical process behind this switch between one regime or another can be made by the following reasoning. In astrophysical conditions the classical resistivity is very small, and the magnetic field behaves macroscopically as if the diffusion term in the magnetic induction equation would be zero. This behaviour is controlled by the magnetic Reynolds number $R_m$,
\begin{equation}
R_m = \frac{UL}{\eta },
 \end{equation}
where $U$ and $L$ are a characteristic velocity and a space-scale, respectively.  From a physical point of view $R_m$ is a measure of the size of the advection term, $\nabla \times \left(\vec{v}\times \vec{B}\right)$, as compared with the size of the diffusion term, $\eta \Delta \vec{B}$. When $R_m>>1$, the diffusion term in Eq.~(\ref{magnind}) is negligible, while for $R_m<<1$, the evolution of the magnetic field is purely diffusive.

However, under certain conditions, in relatively small volumes, the diffusive behaviour becomes dominant, and the magnetic field lines reconnect. To model this process in our grid, we assume that the astrophysical evolution occurs in two different regimes, with the switch between these two regimes determined by the behaviour of the control parameter.  The control parameter tells us what is the value of the difference between the magnetic field between one grid point, and its neighbours. By definition, this control parameter is thus local, and its characteristic scale $l$ is very small. If $L$ is the characteristic length scale of the simulation, then in our case $l/L$ is at least $0.002$. If the control parameter exceeds a certain threshold, then what happens locally becomes worthwhile inspecting. Since we may assume that the velocity does not change in order of magnitude, and since the diffusivity $\eta$ is constant (the plasma properties do not change), the ratio between the macroscopic Reynolds number and that of the local Reynolds number is of the same order of $l/L$. This can be viewed as a reason why the control parameter changes the behaviour of the Reynolds number.

To summarise, the advective regime is the main framework in which we develop our model. If the control parameter becomes critical the magnetic field evolution is given, for a brief period of time, by a diffusive behaviour. Once this local criticality is relaxed, the control is given back to the advective evolution.

Analytically,
\begin{equation}
\frac{\partial B_x}{\partial t} = \left \{ \begin{array}{rl}  -\frac{\partial (B_xv_z)}{\partial z}, & \text{ high } R_m>>1,  \\ \eta \left(\frac{{{\partial ^2}B_x}}{{\partial {y^2}}} + \frac{{{\partial ^2}B_x}}{{\partial {z^2}}}\right), & \text{ low } R_m<<1.\\
\end{array}\right .
\end{equation}

We take the term $- v_z\frac{{\partial B_x}}{{\partial z}}$ as the stochastic source term and denote it by $S(y,z,t)$.

The evolution equations are brought to dimensionless form by the scalings
\begin{eqnarray}
t&=&\alpha T, B_x = b(T, Z,Y){B_0},z = \beta Z,y = \gamma Y,\nonumber\\
v& =& {v_0}V,G_x = g{B_0}/{\beta ^2},
\end{eqnarray}
where $\alpha $, $B_0$, $\beta $, $\gamma $ and $v_0$ are constants.
With these transformations, the system description becomes

\begin{equation}\frac{{\partial b}}{{\partial T}} = k\frac{{{\partial ^2}b}}{{\partial {z^2}}} + \sigma k\frac{{{\partial ^2}b}}{{\partial {y^2}}},\quad k = \frac{{\alpha \eta }}{{{\beta ^2}}},\sigma  = \frac{{{\beta ^2}}}{{{\gamma ^2}}},\end{equation}
for the diffusive behaviour, and
\begin{equation}\frac{{\partial b}}{{\partial T}} =  - \chi b\frac{{\partial V}}{{\partial Z}} + s(Y,Z,T), \quad \chi  = \frac{{\alpha {v_0}}}{\beta },\end{equation}
for the advective behaviour, where $s(Y,Z,T) = \alpha  S(y,z,t) /B_0$; the dimensionless expression for the control parameter is
\begin{equation}g =  - \frac{1}{4}\left( {\frac{{{\partial ^2}b}}{{\partial {Z^2}}} + \sigma \frac{{{\partial ^2}b}}{{\partial {Y^2}}}} \right).\end{equation}

In the Solar Corona the magnetic diffusivity $\eta $ can be approximated as $\eta \approx 10^4\left(T/10^6\;{\rm K}\right)^{-3/2}$ cm$^2$/s, while the diffusion region has a size of $L\approx 2\times 10^8$ cm \cite{Sol}. By adopting a characteristic velocity of the order of $U=10^3$ cm/s, it turns out that the magnetic Reynolds number of a plasma with temperature $T=10^6 $ K is around $R_m=2\times 10^7>>1$. In this case the diffusion term in the magnetic diffusion equation Eq.~(\ref{magnind}) is negligible. However, in Solar Physics there are several important exceptions, like, for example, in the neighbourhood of magnetic neutral points and lines, during magnetic reconnection, and during solar flares, when the diffusion processes play an important role in the understanding of the corresponding processes \cite{Sol}. In the case of the Gamma Ray Bursts, one can estimate the magnetic Reynolds number by adopting for the magnetic diffusion coefficient $\eta $  the perpendicular
resistivity in a strong magnetic field $\eta _{\perp}=1.3\times 10^{13} \times Z\ln \Lambda /T^{3/2}$ cm$^2$/s, where $\Lambda = 3/2e^3
\sqrt{k^3T^ 3/\pi n}$ is the Coulomb logarithm, with $k$ denoting Boltzmann's constant and $n$ the particle number density, respectively \cite{Zhang}. By assuming physical conditions specific for GRBs, we find $\eta _{\perp}\approx 1$ cm$^2$/s, leading to a large magnetic Reynolds number of the order of $R_m\approx 10^{24}$. However, by assuming for the magnetic diffusivity in GRB magnetized plasma its maximum value $\eta _B$, corresponding to the Bohm diffusion, we obtain $\eta _B\approx r_Bc=\gamma _em_ec^3/eB$, where $r_B=\gamma _em_ec^2/eB$ is the co-moving frame cyclotron radius, and $\gamma _e$ is the electron Lorentz factor. Hence for the magnetic Reynolds number in the Bohm diffusion approximation $R_{m,{\rm Bohm}}$ we obtain $R_{m,{\rm Bohm}}\approx 3.4\times 10^{12}\gamma _e^{-1}L_{w,52}^{1/2}\Gamma _{2.5}^{-2}\sqrt{\sigma /(1+\sigma)}$ cm$^2$/s, where $L_{w,52}$ is the luminosity of the GRB ejecta in units of $10^{52}$ erg/s, $\Gamma _{2.5} $ is Lorentz factor of the ejecta in units of $10^{2.5}$, and $\sigma $ is the magnetization parameter \cite{Zhang}. Hence for high values of the particle and ejected wind Lorentz factors, and for relatively low  Gamma Ray Burst luminosities, in the Bohm diffusion limit considerably small magnetic Reynolds numbers may characterize the plasma ejected by the GRB explosion. For example, for $\gamma _e=10^{10}$, $L_{w,52}=1$, $\Gamma _{2.5}=100$, and $\sigma >>1$, $R_{m,{\rm Bohm}}\approx 3.4\times 10^{-2}$. Therefore, during the post-explosion expansion of the relativistic fireball in a strong magnetic field both very high and low magnetic Reynolds number regimes may be present, and this possibility must be taken into account when analysing the dynamical behaviour of the GRB lightcurves.

\subsection{The discretized model}

Following the regular discretisation procedure (with indices $i,j,k$ standing for the $Y$, $Z$ and $T$ coordinates), we obtain for the above equations
\begin{eqnarray}
\frac{{{b_{i,j,k + 1}} - {b_{i,j,k}}}}{{\Delta T}} = \frac{k}{{{{\left( {\Delta Z} \right)}^2}}}\left[ {{b_{i,j + 1,k}} + {b_{i,j - 1,k}} - 2{b_{i,j,k}}} \right] \\ \nonumber + \frac{{k\sigma }}{{{{\left( {\Delta Y} \right)}^2}}}\left[ {{b_{i + 1,j,k}} + {b_{i - 1,j,k}} - 2{b_{i,j,k}}} \right],
\end{eqnarray}
for the diffusive behaviour,
\begin{equation}\frac{{{b_{i,j,k + 1}} - {b_{i,j,k}}}}{{\Delta T}} =  - \chi {b_{i,j,k}}\frac{{{V_{i,j + 1,k}} - {V_{i,j,k}}}}{{\Delta Z}} + {s_{i,j,k}},\end{equation}
for the advective behaviour and
\begin{eqnarray}{g_{i,j,k}} =  - \frac{1}{4}\frac{1}{{{{\left( {\Delta Z} \right)}^2}}}\left[ {{b_{i,j + 1,k}} + {b_{i,j - 1,k}} - 2{b_{i,j,k}}} \right] \\ \nonumber- \frac{1}{4}\frac{\sigma }{{{{\left( {\Delta Y} \right)}^2}}}\left[ {{b_{i + 1,j,k}} + {b_{i - 1,j,k}} - 2{b_{i,j,k}}} \right],\label{eq:critPar}
\end{eqnarray}
for the critical parameter.

We will consider flows with a velocity decreasing as the spatial grid index increases, such that

\begin{equation}
V_{i,j+1,k}<V_{i,j,k},
\end{equation}
and, even more, propagating flows such that the velocity is zero for points not yet reached by the wave-front along the vertical direction $Z$, i.e.,
\begin{equation}
V_{i,j+1,k}-V_{i,j,k} = -V_{i,j,k}.
\end{equation}
For the behaviour of the wave-front with respect to the $Y$ axis we consider a Gaussian function centred on the current value of the $Z$ coordinate, of a width fixed in the code, but such that the integral of the Gaussian with respect to $Y$ is $1$.

When the critical threshold has been reached in a point, the diffusive behaviour of the induction equation is
\begin{equation}{b_{i,j,k + 1}} \to {b_{i,j,k}} - \frac{4}{5}{g_{i,j,k}}\label{eq:point}\end{equation}
and the redistribution is
\begin{equation}{b_{i,j \pm 1,k + 1}} \to {b_{i,j,k}} + \frac{1}{5}{g_{i,j,k}}\label{eq:neigh1},\end{equation}
\begin{equation}{b_{i \pm 1,j,k + 1}} \to {b_{i,j,k}} + \frac{\sigma }{5}{g_{i,j,k}}.\label{eq:neigh2}\end{equation}
The energy released by each volume during an individual relaxation event is the magnetic energy lost in that volume
\begin{equation}{E_R} = \frac{1}{{2{\mu _0}}}\int_{dz} {\int_{dy} {\int_{dx} {\left[ {{B^2}_{x\;(in)} - {B^2}_{x\;(out)}} \right]dx} dy} dz} \end{equation}
and in dimensionless form it is
\begin{equation}{e_R} = \frac{1}{{2{\mu _0}}}\int_{dZ} {\int_{dY} {\left[ {{B^2}_{x\;(in)} - {B^2}_{x\;(out)}} \right]dY} dZ}, \end{equation}
where
\begin{equation}{E_R} = {e_R}\frac{{\beta \gamma B_0^2}}{{2{\mu _0}}}.\end{equation}
Hence we find
\begin{equation}{e_k} = \sum\limits_i {\sum\limits_j {\left( {\frac{8}{5}{b_{i,j,k*}}{g_{i,j,k*}} - \frac{{16}}{{25}}g_{i,j,k*}^2} \right)} } .\end{equation}

The star on the timestep counter $k$ represents the fact that within one time step $k$ of the simulation, the same
cell, due to next neighbour interaction might become unstable
more than once. The $k*$ is a subdivision of the simulation
time step and it is non-zero as long as the cell is unstable.

\subsection{The simulation procedure}

The simulation procedure is described below as follows:
\begin{enumerate}
\item{} Initialization: a two dimensional grid with $N_Y \times N_Z$ cells is initialized in each cell with the value $b_0$ ; the initial flow velocity $(V_{0,0,0})$ is some multiple of the characteristic Alfven speed for the configuration, given by $\chi$;
\item{} Evolution: for each $k^{th}$ timestep in the interval $\overline{1,N_T}$, the evolution of the system is as follows:
    \begin{itemize}
    \item{} Since the upward flow with velocity $V$ is deterministic, one can formally know what cell $j_k$ the flow has reached at the time step $k$. For fixed current $Z$ coordinate, the $Y$ cells associated to it are updated with values drawn from a Gaussian centred on the current $Z$, with standard deviation $N_Y/20$ (advective behaviour);
    \item{} A random pair of numbers \{$k_i$, $k_j$\}, $k_i\in \overline{1,N_Y}$, $k_j\in \overline{1,N_Z}$ is chosen and updated as
    \begin{equation}
    s_{i,j,k} = b_{i,j,k}(1+\Delta T \epsilon),
    \end{equation}
    where $\epsilon <1$ is a positive small number (stochastic loading);
    \item{} The mean value of the magnetic field $b$ is calculated; each component of the grid is then scaled with respect to this mean value. The critical parameter $g_{cr}$ is taken as 10\% of this value.
    \item{} Scanning: The control parameter is calculated with Equation~\eqref{eq:critPar} for each cell in the grid and stored into a matrix. If any one cell has an absolute value higher than the control parameter $g_{cr}$ a flag is triggered;
    \item{}Redistribution: If the flag has been triggered, the control parameter matrix will be searched for positive values that are higher than $g_{cr}$. If any are found, part of the cell's content will be redistributed to its nearest neighbours according to Eqs.~\eqref{eq:point}-\eqref{eq:neigh2}. If no positive values were found, the matrix will be searched for negative values with absolute value higher than $g_{cr}$. This sweep is done while cells with value higher than the critical parameter are found. A variable $N_k$ stores the number of such events for each step $k$. Another variable $S_k$ stores the number of cells reached by the critical flow.
     \end{itemize}
\item{} Results: $N_k$ represents the number of events needed to fully relax the grid at each time step $k$; the vector $N$ is used to produce the lifetime distribution, $D(N)$; similarly, $S_k$ represents the number of cells reached by the critical flow at each time step $k$; the vector $S_k$ is used to produce the lifetime distribution, $p(S)$.
\end{enumerate}

The connection between grid parameters, observational parameters and simulation output is given in Table~\ref{tab:input}.

\begin{figure*}
   \includegraphics[width=0.5\columnwidth]{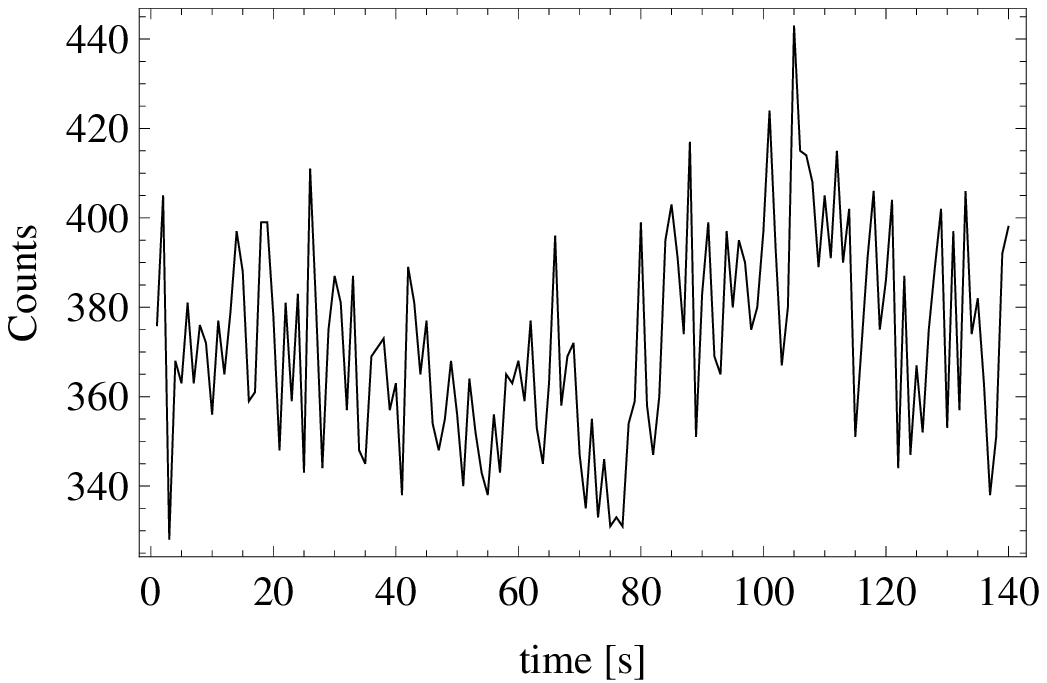}
   \includegraphics[width=0.5\columnwidth]{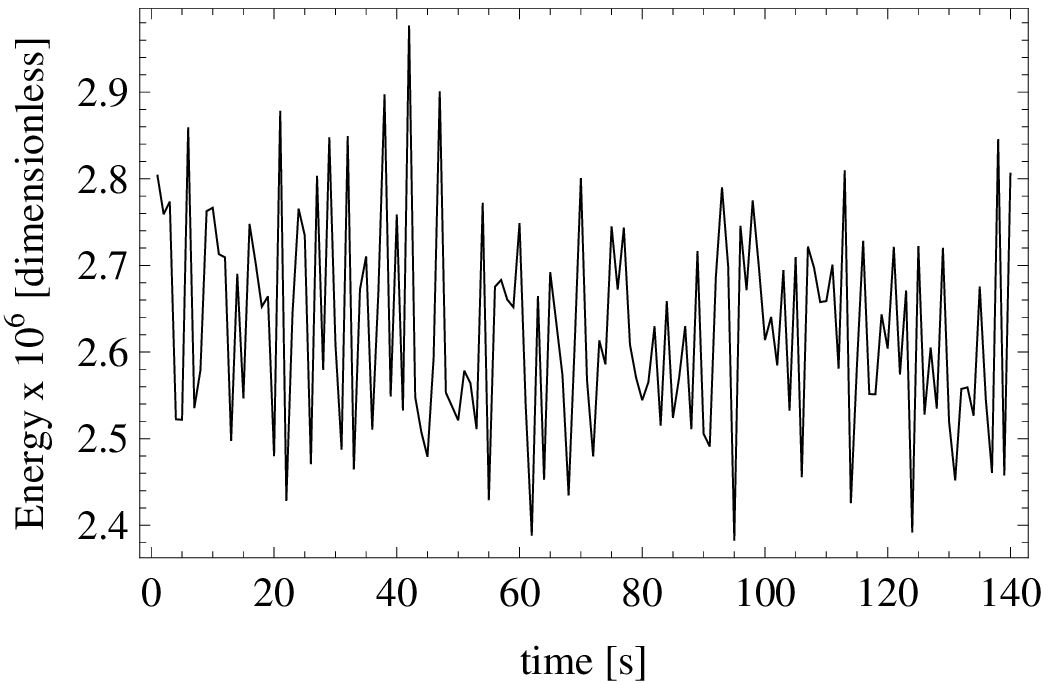}
   \caption{Left panel: plot of the Counts vs. time data of GRB 140919636 on 2014-09-19, obtained with the FERMI Gamma-ray Space Telescope with detector 0 \cite{Dat}. Right panel: Simulated energy release as a function of time, for $\chi =50$ and the $V_k = $const. velocity profile, caused by a single GRB pulse impinging at the base of the simulation grid.}
   \label{fig:simObs1}
  \end{figure*}

     \begin{figure*}
   \includegraphics[width=0.5\columnwidth]{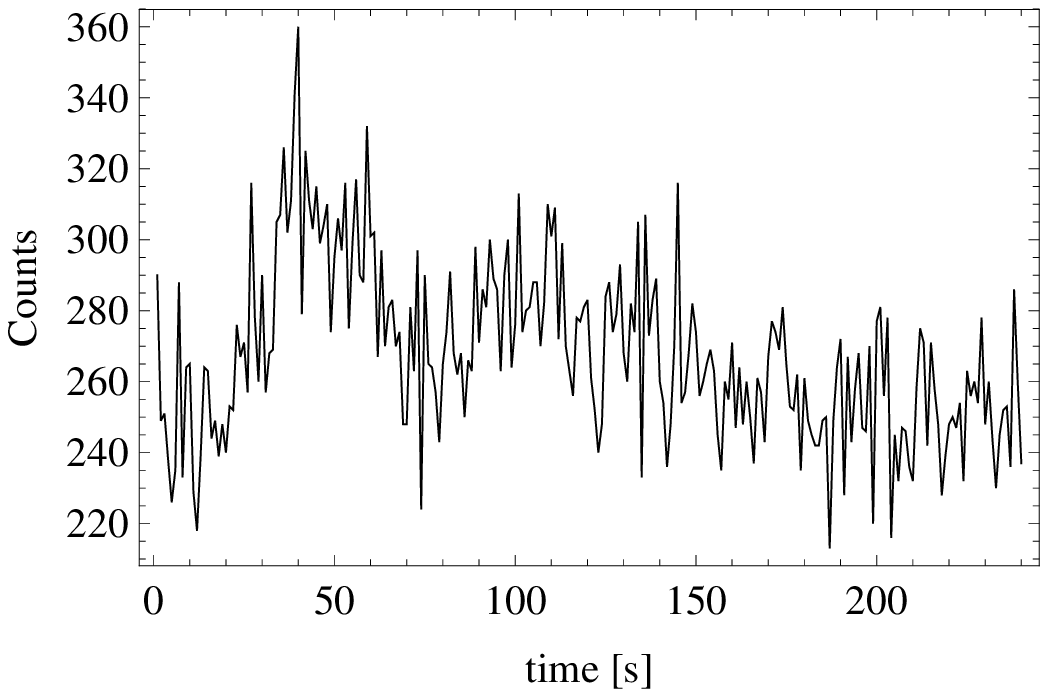}
   \includegraphics[width=0.5\columnwidth]{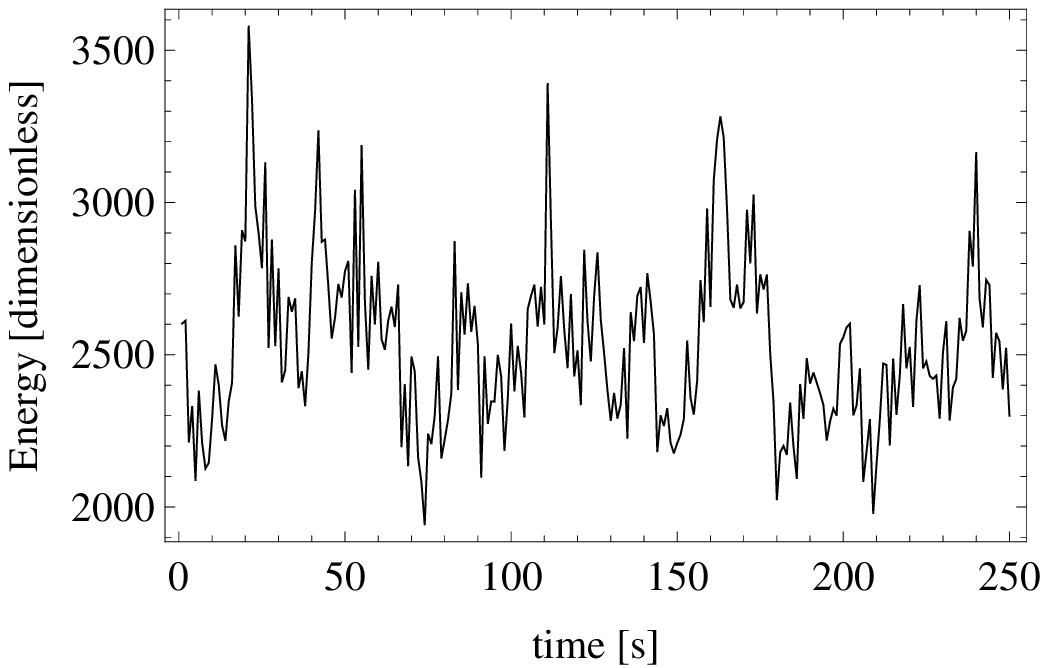}
   \caption{Left panel: plot of the Counts vs. time data of GRB 080805496 on 2008-08-05 obtained with the FERMI Gamma-ray Space Telescope with detector 0  \cite{Dat1}. Right panel: Simulated energy release as a function of time, for $\chi =10$ and the $V_k = $const. velocity profile, caused by a single GRB pulse impinging at the base of the simulation grid.}
   \label{fig:simObs3}
  \end{figure*}

   \begin{figure*}
   \includegraphics[width=0.5\columnwidth]{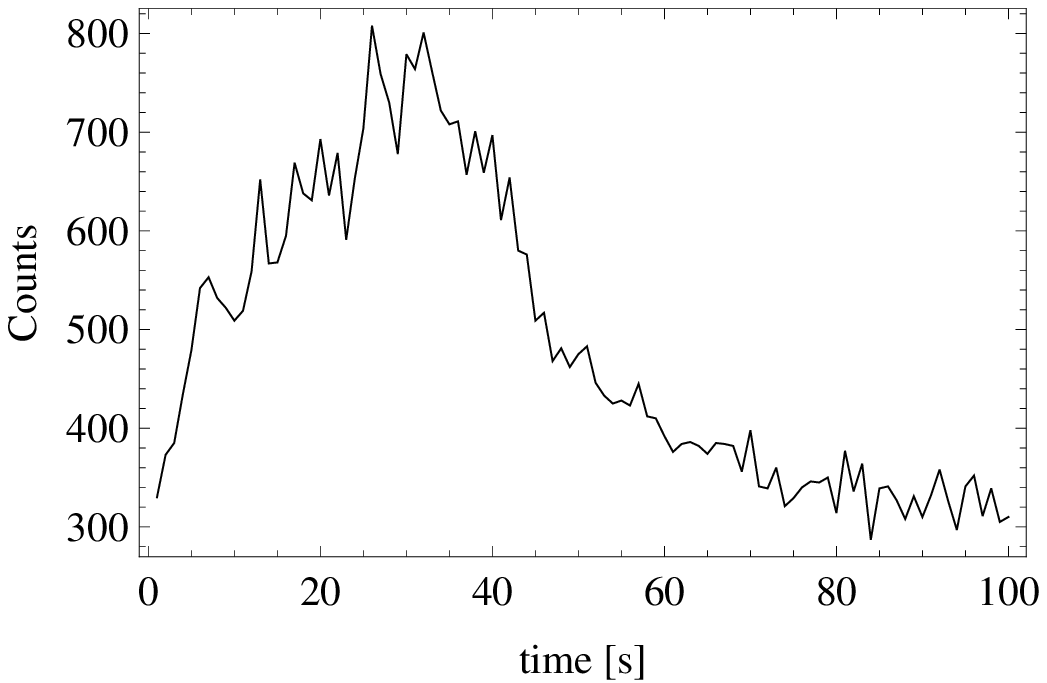}
   \includegraphics[width=0.5\columnwidth]{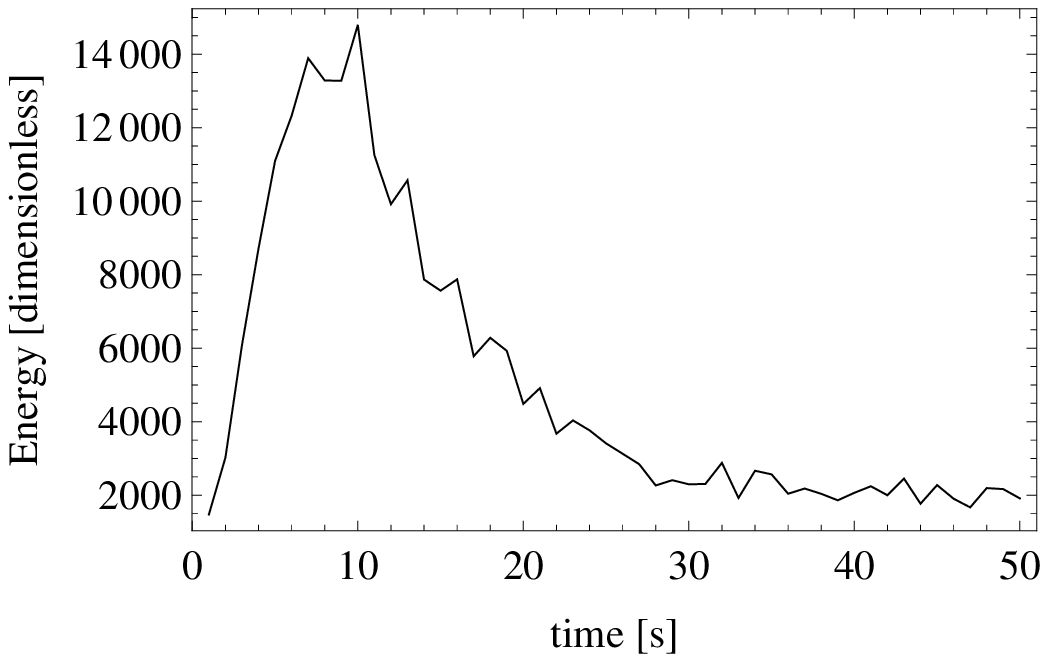}
   \caption{Left panel: plot of the Counts vs time data of GRB 101126198 on 2010-11-26, obtained with the FERMI Gamma-ray Space Telescope with detector 7 \cite{Dat2} . Right panel: Energy release as a function of time, for $\chi =10$ and the $1/\sqrt{k} $ velocity profile, caused by a single GRB pulse impinging at the base of the simulation grid.}
   \label{fig:simObs8}
  \end{figure*}

\begin{table}\caption{Parameter correlations. The Dimensionless column contains parameters which are set beforehand, and which generally characterise the simulation grid; the Independent parameters are those set by observations; the Dependent column contains the parameters with an analytical dependency with respect to the dimensionless and/or independent parameters.}
\begin{tabular}
{| p{0.35\linewidth}  | p{0.27\linewidth} | p{0.27\linewidth}|}\hline
   Dimensionless 					 	& Dependent 							& Independent		\\ \hline
    $N_T=10^5$ 					     	& $\beta = \alpha v_A$					& $B_0=10^{14}\;{\rm G}$ (observations)	\\ \hline
    $N_Z=500$  	
    $N_Y=50$					 	    & $E_R \sim \frac{\beta ^2 B_0^2}{2\mu \sqrt{\sigma}}$	& $v_A = 10^9\;{\rm cm/s}$ (corresponding to $B_0$)\\ \hline
    $\Delta T = \Delta Z = \Delta Y = 1$& 	& \\ \hline
    $\chi \in (0,1,5,10,50,100)$  	& 	    & \\ \hline

   	$\epsilon = 0.3$				    &          &  \\ \hline
   	$\sigma = 1$					    & 		&\\ \hline

\end{tabular}\label{tab:input}
\end{table}

In order to obtain a description of the stochastic behaviour of the light curves we use the slope of the Power Spectral
Distribution (PSD) of the luminosity.  The slopes of the PSD curves can provide some important insights into the nature of the physical mechanisms leading to the observed variability of the astrophysical source. If $X$ is a fluctuating
stationary physical quantity, with mean $\mu _X$ and variance $\sigma _X ^2$, respectively, then we define
the autocorrelation function for  $X$ as follows~\cite{Vaseghi}
\begin{equation}
R_X(\tau) = \frac{\langle \left ( X_s - \mu _X \right ) \left (
X_{s+\tau} - \mu \right) \rangle}{\sigma _X^2},
\end{equation}
where $X_s$ denote the numerical values of $X$ measured at time $s$, and $\langle
\rangle$ denotes averaging over all values $s$, respectively. Then the PSD is defined as~\cite{Vaseghi}
\begin{equation}
P(f) = \int _{-\infty} ^{+\infty} R_X(\tau) e^{-\imath 2\pi f \tau}
d\tau.
\end{equation}

From a physical point of view the slope of the PSD of a time series $X$ provides some statistical insight to the degree of correlation
physical processes have with themselves. The observational light curves of the GRBs, as well as the simulated ones, are analysed in the next Section.

\section{Results and astrophysical implications}\label{sect2}

We ran simulations in which we varied the time profile of the dimensionless velocity $V_{k}$ according to the laws
\begin{equation}
V_{k} = {\rm constant}, V_{k} \sim \sqrt{k^{-1}}, V_{k} \sim k^{-1},
\end{equation}
and the value of $\chi$. These velocity profiles were chosen to mimic an ejected plasma flow placed in a gravitational field for a few specific cases: for $V_{k} = {\rm constant}$ the ejecta is so energetic that it does not feel any gravitational pull and the initial value of the velocity is conserved; the profiles $V_{k} \sim \sqrt{k^{-1}}$ and $V_{k} \sim k^{-1}$ are implemented to take into account a temporal decrease of velocity (equivalently, the plasma velocity decreases as a function of distance to the expulsion site).

\begin{figure}
   \includegraphics[width=0.5\columnwidth]{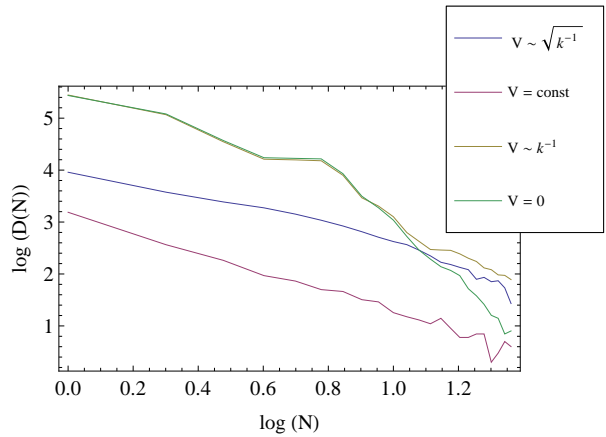}
   \caption{The lifetime distribution $\log(D)$ is plotted against the number of events $\log(N)$, for $\chi = 100$ and different velocity profiles.}
   \label{fig:D(N)Chi100}
   \end{figure}

\begin{figure}
   \includegraphics[width=0.5\columnwidth]{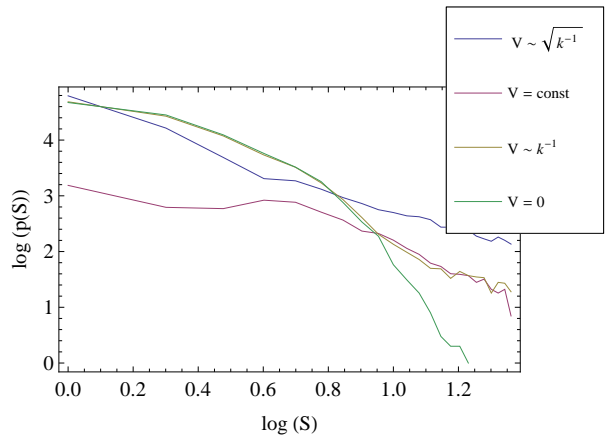}
   \caption{The size distribution $\log(p(S))$ is plotted against the size of the avalanche $\log(S)$, for $\chi = 100$ and different velocity profiles.}
   \label{fig:SizeChi100}
   \end{figure}

\begin{figure}
   \includegraphics[width=0.5\columnwidth]{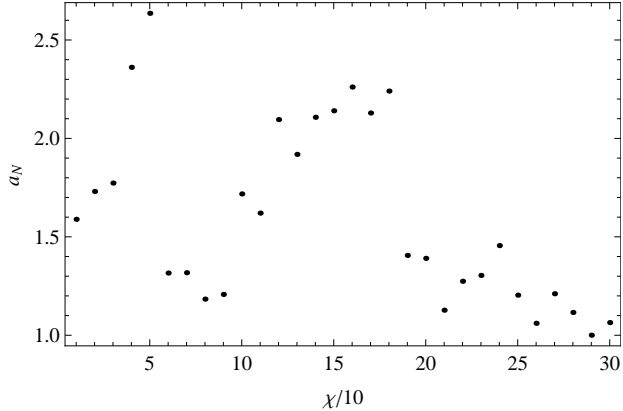}
   \caption{A fit of the type $D(N)\sim N^{-a_N}$ is made for the lifetime distribution. The plot shows the values of $a_N$ as a function of $\chi$, for $V \sim \sqrt{k ^{-1}}$.}
   \label{fig:slopeVsChi}
  \end{figure}

\begin{figure}
   \includegraphics[width=0.5\columnwidth]{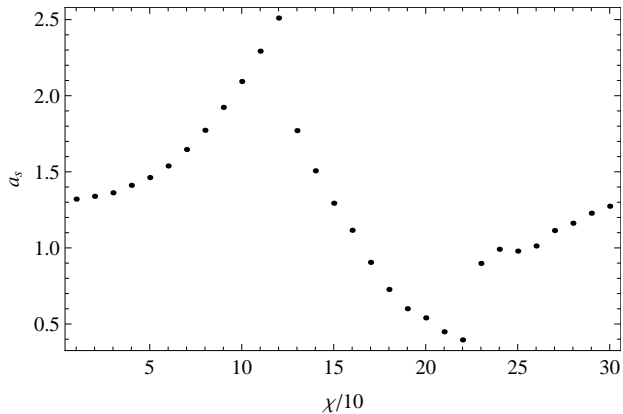}
   \caption{A fit of the type $p(S)\sim S^{-a_S}$ is made for the event size distribution. The plot shows the values of $a_S$ as a function of $\chi$, for $V \sim \sqrt{k ^{-1}}$.}
   \label{fig:slopeVsChi-size}
  \end{figure}

The main results of the simulations are the timestep evolution of the control parameter, the timestep evolution of the energy release, represented in the right panels of Figs.~\ref{fig:simObs1} - \ref{fig:simObs8}, the event lifetime distribution (Fig~\ref{fig:D(N)Chi100}), the event size distribution (Fig.~\ref{fig:SizeChi100}) and the values of the spectral slope of the event lifetime distribution (Fig.~\ref{fig:slopeVsChi}) and of the event size distribution (Fig.~\ref{fig:slopeVsChi-size}), for different values of $\chi$.

To allow a quick comparison with the observational data, in the left panels of Figs.~\ref{fig:simObs1} - \ref{fig:simObs8} we present the light curves (counts) for three GRBs, with the plots done with the data file supplied by NASA's HEASARC Data access database \cite{Dat1,Dat2,Dat}.

We identify the number of events needed to relax one critical onset with the lifetime of this avalanche. Thus, the distribution for lifetimes is equivalent to the distribution of number of events,
\begin{equation}
D(N)\sim N^{-a_N}.
\end{equation}
For low advective velocities ($\chi \sim 1$) the exponent $a_N$ of a power-law distribution is approximately $1.6$, which is in agreement with the theoretical value for two dimensional grids (see e.g. Eq. (3.7) from \cite{bak1988}). It is customary to argue that the exponent $a_S$ of the power-law fitting the event size distribution is equal to the dimensionality of the system (i.e., $a_S = 2$ for a two-dimensional system) \cite{Wang}. This is based on an unbiased diffusive random walk argument, where the distance $L$ reached by a random walker in time $T$ is $L\sim \sqrt{T}$. More precisely, the walker is equally likely to choose any direction for his next step. This is not the case in our work, due to the inherent asymmetries of a deterministic background flow. As such, the numerical values of $a_S$ depart from the value of $2$.

\begin{figure}
   \includegraphics[width=0.5\columnwidth]{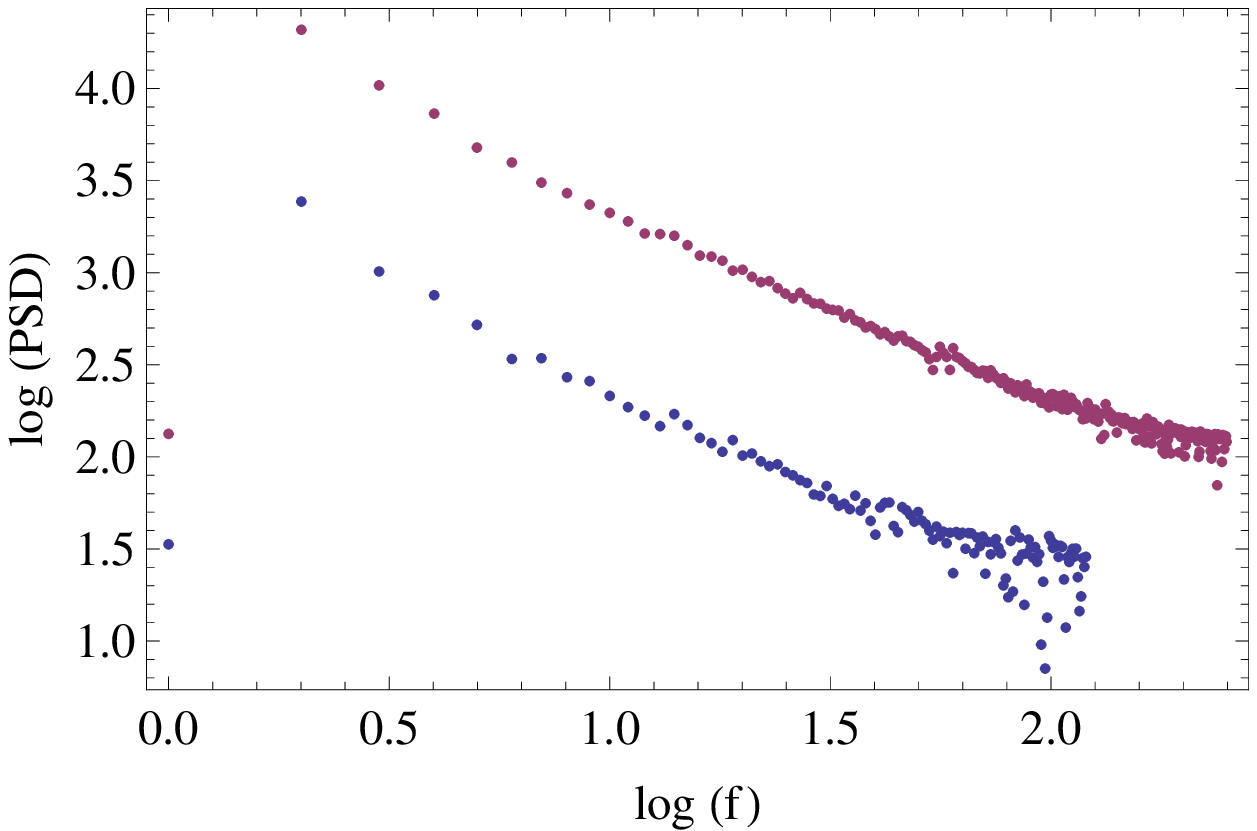}
   \caption{Comparison of the variation of $\log(PSD)$ as a function of $\log (f)$  for the light curve of GRB GRB100919884 (upper curve) and for the
simulated light curve with $V_k\sim 1/\sqrt{k}$ and $\chi =10$ (lower curve).}
   \label{fig:compPsd}
   \end{figure}

In order to produce a more quantitative comparison, we calculate the PSD of both the observed and simulated light curves; this is done by taking the corresponding  time series, calculating their correlation functions, and then taking the Fourier transform. We then fit a model of the type $PSD (f) \sim f^{-\alpha}$ to them. The results from this fit, shown in Table~\ref{tab:ObsPSD}, agree well with the results from the simulations as presented in Table~\ref{tab:SimPSD}. For illustration purposes, the PSD for GRB100919884 is shown on the same plot with the PSD of a simulated light curve in Fig.~\ref{fig:compPsd}. The $R^2$ statistics presented in the third column of both these Tables is obtained following a linear regression scheme, based on minimizing the sum of the squared residuals; it is calculated as the ratio of the ratio of the model sum of squares to the total sum of squares.

\begin{table}\caption{PSD of observations; a linear fit $\log{PSD(f)} = -\alpha \log{f}$ was done. The third column shows the value of the $R^2$ statistics for the linear fit.}
\begin{tabular}
{| p{0.35\linewidth}  | p{0.27\linewidth} | p{0.27\linewidth}|}\hline
   Data identification					 	& $\alpha$ 							& $R^2$		\\ \hline
    GRB140919636 & 0.708 & 0.806 \\ \hline
    GRB141223240 & 0.585 & 0.775 \\ \hline
    GRB080805496 & 0.746 & 0.804 \\ \hline
    GRB130702004 & 0.960 & 0.832 \\ \hline
    GRB120119170 & 1.505 & 0.881 \\ \hline
    GRB100919884 & 0.912 & 0.810 \\ \hline
    GRB131230808 & 0.722 & 0.788 \\ \hline
    GRB101126198 & 1.698 & 0.881 \\ \hline
    GRB091209001 & 1.006 & 0.827 \\ \hline
\end{tabular}\label{tab:ObsPSD}
\end{table}

\begin{table}\caption{PSD of simulations; a linear fit $\log{PSD(f)} = -\alpha \log{f}$ was done. The third column shows the value of the $R^2$ statistics for the linear fit.}
\begin{tabular}
{| p{0.35\linewidth}  | p{0.27\linewidth} | p{0.27\linewidth}|}\hline
   Setup identification					 	& $\alpha$ 							& $R^2$		\\ \hline
    $V_k\sim 1/k$, $\chi = 1$ & 1.125 & 0.877 \\ \hline
    $V_k\sim 1/k$, $\chi = 50$ & 1.119 & 0.890 \\ \hline
    $V_k\sim 1/\sqrt{k}$, $\chi = 1$ & 1.160 & 0.880 \\ \hline
    $V_k\sim 1/\sqrt{k}$, $\chi = 10$ & 1.118 & 0.876 \\ \hline
    $V_k = const$, $\chi = 10$ & 1.703 & 0.915\\ \hline
\end{tabular}\label{tab:SimPSD}
\end{table}

One important problem in CA-SOC simulations is the connection (conversion factor) between the timestep in the simulations and real time. This problem is even greater in simulations such as ours, which include two timescales: the advection timescale, set by the upwards propagating shock wave and the SOC timescale, i.e., the time needed by an avalanche to take place.

As we see it, the main problem here is that, after SOC sets in, one timestep is the time needed for an avalanche to occur and this is not a constant. But in order to really be able to compare between observations and simulations, a connection between real lapsed time and simulation timestep is proposed as follows: the magnetic reconnection rate in the Sun is of the order of $10^{-2}-10^{-3}$~\cite{isobe,nagashimi}; reconnection rates in various astrophysical contexts do not vary that much, and hence we assume that this is also the order of magnitude of the reconnection in the GRB flow; we thus set $\alpha = 10^{-4}$; we do this because $\alpha$ needs to be in a relationship to the smallest timescale in the problem such that all possible time-features can be probed by the simulation, i.e., $\alpha \ll \tau _{reconnection}$.

Next, we assume that once reconnection begins, the changing field line stresses will be transmitted throughout the simulation domain at the Alfven speed~\cite{lazarian}. Thus, the reconnection time is equal to the Alfven time, which is $10^{-4}\;{\rm s}$ for our case (calculated using Table~\ref{tab:input}). Then we assume that one timestep can be converted to seconds by multiplying the duration of a reconnection event  by the average number of events produced by the simulation, estimated to be around $10$ (see Figure~\ref{fig:D(N)Chi100}). In the end, we conclude that $10^3$ timesteps make for one real second of observation time.

In order to estimate the energy in dimensions of $erg$ produced by the simulation we use Table~\ref{tab:input} to get the order of magnitude for an $E_r$ produced by one timestep, $E_r^{timestep}\approx 4 \cdot 10^{38} {\rm erg}$; if we assume that the conversion factor from time step to second is $10^3$, than the conversion factor between the dimensionless energy and the physical real energy is approximately $ 4 \times 10^{41} {\rm erg}$. The observed energy output in GRB explosions is around $10^{44} {\rm erg}$ ~\cite{Chin}; the interval of dimensionless value for the energy obtained by simulations covers the values needed to produce agreement with observations.

\begin{table*}
 \centering
 \begin{minipage}{140mm}\caption{The first column contains valid inquiries about observations, simulations and theory and how and if they agree. The second column contains an assessment of whether or not it is suitable to compare the simulation output with observational data. The third and fourth column contain the results of such comparison with observations and theory respectively. We use a $-$ mark when such a comparison cannot be made, a $\checkmark$ mark when the comparison produces good results and a $\times$ when the comparison fails.}
\begin{tabular}
{| p{0.35\linewidth}  | p{0.45\linewidth} |p{0.10\linewidth} | p{0.10\linewidth} |} \hline
  Output of the simulation & May be compared with observational GRB data & Obs. & Theory \\ \hline
  Event lifetime distribution (analytical dependency $N(E) \sim N^{-a}$) & yes & $\checkmark$ & $\checkmark$ \\ \hline
  PSD (analytical dependency $P(f) \sim f^{-\alpha}$) & yes & $\checkmark$ & $\checkmark$ \\ \hline
  Light curve & yes, a qualitative visual comparison & $\checkmark$ & - \\ \hline
  Maximum released energy & yes, if the proposed conversion factor from dimensionless to dimensional is accepted & $\checkmark$ & - \\ \hline
  Number of peaks in the light curve & yes, for some types of data & $\checkmark$ & - \\ \hline
  Waiting time in the LC  &  yes, if the waiting time is defined as time between subsequent avalanches in the grid (SOC characteristic) \newline no, if waiting time is defined as lapsed time between initial pulse and maximum of the LC, because depends on the initial condition in the grid which is unknown \newline no, if waiting time is defined as time lapsed between independent prominent light curve features, i.e. as those caused by different pulses at the base & - & $\checkmark$ \\ \hline
  Critical parameter timestep evolution & no, this is just for SOC validation. For the moment this is just a theoretical tool. & - & $\checkmark$ \\ \hline
  Critical parameter maximum value & no, although this could be done in principle if magnetograms of the studied regions would be available& - & -\\ \hline
  Magnetic field divergence & no, this is just for general model validation (fundamental physics constraint) &- & $\checkmark$\\ \hline

\end{tabular}\label{tab:output}
\end{minipage}
\end{table*}

We note one very interesting fact: {\it for zero velocity (i.e., no pulse at the base of the simulation grid), the model produces an event lifetime distribution slope in agreement with theory.} For low impulse velocity, the event lifetime distribution slope is still close to the theoretic one. However, when $\chi$ increases, for certain velocity profiles, the slope of $D(N)$ vs. $N$ departs from its theoretical value. The issue is best illustrated by plotting the value of the spectral slope as a function of $\chi$ (Fig.~\ref{fig:slopeVsChi}). It is clear that for high values of $\chi$, one obtains a spectral slope with a value very close to $1$, the theoretical accepted spectral slope for 1D SOC. {\it From a strict theoretical point of view, one dimensional magnetic reconnection is not possible; it was thus a problem (swept under the rug) to say that while the observational data for GRBs shows that a one dimensional SOC is occurring, this is due to magnetic reconnection.}

{\it However, our results can be interpreted as showing that the physics is genuinely 2D, but an energetic initial impulse leads to a 1D SOC signature.}

The physics and observations of GRBs are extremely complex. It is understandable that a simple CA model cannot and does not reproduce all features of the real process. As such, we list some of the caveats of our model: it does not produce information about bulk Lorentz factor, about isotropy (or lack thereof) of the radiation. Also, although under some assumptions, magnetic CA models can produce maps of the magnetic field on the microscopic level, this was not the purpose of this paper. A discussion of how well the model agrees with observational data is presented in Table~\ref{tab:output}.

\section{Conclusions}\label{sect3}

In the present paper we have presented a two dimensional CA model for a magnetized grid with an initial large flow and a stochastic perturbation. The discretized laws for the evolution of the magnetic field were deduced and implemented. This study extends to two dimensions the one dimensional SOC approach introduced in \cite{1dgrb} for the analysis of one dimensional magnetized plasma flows. As a main result of our investigation we have found that the two dimensional magnetized plasma with a background flow system exhibits SOC.  The event lifetime and event size distributions and the emitted electromagnetic energy were obtained, and their astrophysical interpretation was briefly discussed. It was shown that this approach can be used to model GRBs and some of their properties, as summarized in Table~\ref{tab:output}. Most importantly, the model produces results which agree with the observations on the same time and energy scales, which to our knowledge is a novelty.
\cite{Wang} suggested  to use CA and SOC methods to study and interpret the X-ray flares of GRBs. Here we further postulated that the main burst, i.e., the GRB itself, also arise from SOC processes. In contrast to the standard GRB models, where each pulse correspond to one energy injection (e.g., internal shock model), here we show that in the SOC model one energy input is able to generate a series of pulses, which are characteristic of many GRBs. Moreover, we have found that some general features of GRBs (and especially the light curves) can be described in the framework of the present SOC model.

It would be interesting to compare our mathematical SOC formalism with the one used to study the properties of the accretion by black holes. In the model introduced in \cite{Min1,Min2}, the accretion disk is divided into two parts, the outer disk region, and the inner disk region, respectively. In the outer disk region, the gas drifts smoothly inward, while in the inner disk region blobs are formed. The whole disk is divided into 64 rings, and angularly divided into 64 equal parts. The physical quantity stored in the cells is the mass in that region. Since the mass is continuously flowing to the outermost region of the disk, a randomly chosen cell in the ring receives a fixed amount of mass $m$. If the mass of the cell is larger than the critical value $M_{crit}$, a mass flow will occur to the three nearest cells in the adjacent inner ring. By denoting the mass in the cell with coordinates $i$ (representing the radial position) and $j$ (representing the angular position) by $M_{i,j}$, when $M_{i,j}>M_{crit}$, the following CA process happens: $M_{i,j}\rightarrow M_{i,j}-3m$, $M_{i-1,j}\rightarrow M_{i-1,j}+m$, $M_{i-1,j\pm 1}\rightarrow M_{i-1,j\pm 1}+m$. The total X-ray luminosity is of the order of the gravitational potential energy lost in the process, and it can be obtained as $L\approx \sum \left(GMm/r_{i-1}-GMm/r_i\right)$.

In contrast to the above approach, in the present paper we start from the basic result that diffusion type partial differential equations driven by noise can produce characteristic effects similar to SOC \cite{Sor}. Hence, we consider a discretized version of the magnetic induction equation, in the presence of a stochastic term generated by the matter flow and the gradient of the magnetic field. The magnetic induction equation is considered in two limiting cases, corresponding to the diffusive and advective regimes, respectively. There are also significant differences in the physical approach of the two models. While in the accretion disk model of \cite{Min1,Min2} the parameter stored in the grid is the mass, in the present approach the stored quantity is the value of the magnetic field. The disk luminosity is assumed to be purely of gravitational origin in the accretion disk CA models, while in the present case the radiation emission is purely electromagnetic.

An important characteristic of systems that reached SOC is that a slight change in the external conditions of the system
 determines  the generation of avalanches of various sizes.  These avalanches keep, on average, the system near
the critical state \cite{CAmodelling, Gil}. It is important to note that a minor effect can
result in an avalanche of a much bigger size (catastrophe). This kind of catastrophic behaviour could be also responsible for the GRBs explosions, and overall evolution. Therefore SOC and CA methods can be used, and may prove useful, in modelling early phases - explosive or immediately post-explosive - of the GRBs. Once a given critical condition is reached,
it is self-maintained, i.e., there is no
need to vary or fine tune its physical and control parameters. This situation is  quite different as compared, for example,
with the critical state in liquid, where two thermodynamic parameters(the temperature and the density) must be tuned to keep the fluid system
near its critical point.

It is a general property of SOC that the event lifetime and size distributions of avalanches are described in terms
of power laws with negative exponents. This is one of the main reasons why models based on
SOC can be successfully used to describe various natural phenomena. In the present paper we have developed some necessary (but not sufficient) tools that are required for the in depth comparison of the theoretical models with the specific observations of the cosmological Gamma Ray Bursts.

\section*{Acknowledgments}

We thank to the anonymous referee for comments and suggestions that helped us to significantly improve our manuscript. We would like to express our gratitude to Dr.  Pak-Hin Thomas Tam for his help during the preparation of the present work. GM and BD are partially supported by a grant of the Romanian National Authority of Scientific Research, Program for research - Space Technology and Advanced Research - STAR, project number 72/29.11.2013. BD acknowledges the support of ”Babes-Bolyai” University Cluj-Napoca through a Research Excellence scholarship.

\end{document}